# New Scanning Tunneling Microscopy Technique Enables Systematic Study of the Unique Electronic Transition from Graphite to Graphene

P. Xu[a], Yurong Yang[a,b], S.D. Barber[a], J.K. Schoelz[a], D. Qi[a], M.L. Ackerman[a], L. Bellaiche[a], P.M. Thibado[a],*

[a]Department of Physics, University of Arkansas, Fayetteville, Arkansas 72701, USA

[b]Physics Department, Nanjing University of Aeronautics and Astronautics, Nanjing 210016, China

**Abstract**

A series of measurements using a novel technique called electrostatic-manipulation scanning tunneling microscopy were performed on a highly-oriented pyrolytic graphite (HOPG) surface. The electrostatic interaction between the STM tip and the sample can be tuned to produce both reversible and irreversible large-scale vertical movement of the HOPG surface. Under this influence, atomic-resolution STM images reveal that a continuous electronic reconstruction transition from a triangular symmetry, where only alternate atoms are imaged, to a honeycomb structure can be systematically controlled. First-principles calculations reveal that this transition can be related to vertical displacements of the top layer of graphite relative to the bulk. Detailed analysis of the band structure predicts that a transition from parabolic to linear bands occurs after a 0.09 nm displacement of the top layer.

*Corresponding author. E-mail address: thibado@uark.edu (P.M. Thibado)

# 1. Introduction

When a bulk material is cut to form a surface, the broken bonds tend to rearrange into a lower energy configuration in a process known as surface reconstruction. As a result, surface atoms often exhibit a different symmetry than the bulk, such as on the surfaces of Si(001) or GaAs(001) [1, 2]. The atomic arrangement chosen almost always depends on pressure and temperature, and sometimes a particular classification of phase transition can be identified between the various reconstructions [3]. In other cases, a more subtle surface reconstruction occurs, involving only the material's electronic distribution. A prime example is the easily cleaved GaAs(110) surface [4], which exhibits very weak bonding between layers. Therefore when layers are separated, the atomic nuclear positions remain essentially unchanged, but the surface charge density significantly redistributes itself.

A low cleavage-energy system similar to GaAs is graphite. It has long been known that when highly oriented pyrolitic graphite (HOPG) is imaged using scanning tunneling microscopy (STM), only every other atom at the surface contributes to the tunneling current, resulting in an image with trigonal symmetry instead of the expected hexagonal pattern. This is attributed to the particular stacking order most commonly observed in hexagonal graphite [5], referred to as AB or Bernal stacking, wherein half of the surface carbon atoms (the A atoms) are directly above atoms in the layer below, while the other half (the B atoms) are directly above hexagonal holes. The electronic charge density of the A atom is pulled into the bulk, and the STM cannot image it [6]. However, when a single layer of graphite is separated from the bulk, the asymmetry is broken and the subsequent redistribution of the electron density allows every atom to appear in the STM image. This transformation also leads to the other well-known electronic properties that distinguish graphene [7] from graphite, such as a band structure with linear rather than parabolic dispersion in the vicinity of the K-points [8]. Such transitions in the band structure of graphitic layers are especially interesting because they signal that the charge carriers have gained or lost their effective mass, a process of fundamental importance in physics.

Studies using bulk graphite have evidence of graphene; however, the events are randomly occurring. For example, Andrei and coworkers [9, 10] have studied HOPG using STM and low-voltage scanning tunneling spectroscopy (STS). At low temperatures (4.4 K) and after applying a magnetic field, Landau levels consistent with graphene can be observed. Signatures in the sequence have been used to quantitatively predict the amount of interaction between the graphene layer and the bulk. Further evidence of varying degrees of coupling is seen in the symmetry of STM images. The STM tip can provide a perturbation that vertically lifts the top layer [11, 12], resulting in images which exhibit a range of possibilities between the triangular and honeycomb lattices. The difficulty, however, is that this induced decoupling has been mostly random, not lending itself to a systematic study of the important symmetry-breaking transition from bulk graphite to monolayer graphene.

In this article, we present STM images of the HOPG surface before, during, and after perturbing the surface using a new technique we call electrostatic-manipulation STM (EM-STM). With this technique large-scale precision-controlled vertical movement of the HOPG surface is possible. Atomic-scale STM images reveal a continuous transition from graphite to graphene. Density functional theory (DFT) calculations were used to generate a complete set of simulated STM images and provide excellent agreement with the measurements. The continuous change in the spatial distribution of the charge density is proposed as a measure of coupling between the surface layer and bulk.

## 2. Experiments

### 2.1. *STM Measurement details*

The experimental STM images and EM-STM line profiles were obtained using an Omicron ultrahigh-vacuum (base pressure is $10^{-10}$ mbar), low-temperature STM operated at room temperature. The top layers of a 6 mm × 12 mm × 2 mm thick piece of HOPG* were exfoliated with tape to expose a fresh surface. The HOPG was then mounted with silver paint onto a flat tantalum STM sample plate and transferred into the STM chamber, where it was electrically grounded. STM tips were electrochemically etched from 0.25 mm diameter tungsten wire via a custom double lamella setup [13]. After etching, the tips were gently rinsed with distilled water and dipped into a concentrated hydrofluoric acid solution to remove surface oxides [14] before being transferred into the STM chamber. Numerous filled-state STM images of the HOPG surface were acquired using a tip bias of +0.100 V and a constant current of 0.20 nA for small scale images and 1.00 nA for large scale images.

The EM-STM measurements performed were similar in principle to constant-current STS, wherein scanning is paused but the feedback loop controlling the tip's vertical motion remains operational. The STM tip bias is then varied, and one records the vertical displacement required to maintain a constant tunneling current. Assuming the sample is stationary, this process indirectly probes its density of states (DOS). A second interaction is also taking place, though, in which the tip bias induces an image charge in the grounded sample, resulting in an electrostatic attraction that increases with the bias. We have found that in some materials, such as graphite [15] and freestanding graphene [16], this attraction can result in movement of the sample, convoluting and often eclipsing any DOS measurement. In an EM-STM experiment, however, these deformations are actually the subject of interest. By employing electrostatic forces created by the STM tip, one may physically manipulate a surface and examine some of its mechanical properties. Thus an EM-STM measurement involves recording the *z*-position of the tip as the bias is varied at constant current, with the goal of controlled sample manipulation.

---

* ESPI Metals [www.espimetals.com]

## 2.2. *EM-STM on graphite stripe*

The effect of EM-STM on HOPG is demonstrated in Fig. 1. First, a diagram of how this technique might appear on an atomic scale is shown in Fig. 1(a). It illustrates the top layer of HOPG being locally lifted by the electrostatic attraction to the STM tip. A series of 150 nm × 150 nm STM images of HOPG, all at the same location, were taken before, during, and after EM-STM measurements, and the images are displayed in sequential order in Fig. 1(b-f). The slow scan direction proceeded from bottom to top, and the images are colored such that the highest points are white (~2 nm high) while the lowest points are black. A white stripe approximately 20 nm wide is prominent in Fig. 1(b), indicating that a raised ribbon-like structure exists on the HOPG surface. This image was taken prior to any EM-STM measurements. A darker stripe, or trench, can also be seen approximately 50 nm to the right of the white stripe, with a protrusion in the trench serving as a reference point when comparing the images. An EM-STM measurement was taken during the next scan, which is presented in Fig. 1(c). During the EM-STM measurement, the STM tip was first positioned on the white stripe, and then the tip bias was increased from 0.1 V to 3.0 V at a constant tunneling current of 1.00 nA. It can be seen that, at the location where the EM-STM measurements took place, the white stripe was displaced to the right, toward the protrusion, but eventually the upper portion went back to the left, under the influence of the scanning STM tip. In the next image, shown in Fig. 1(d), the lower portion of the white stripe has remained displaced and become somewhat darker (it is likely a fold in the ribbon), indicating that a permanent change has been introduced to the surface. To demonstrate this ability again, a second EM-STM measurement was taken during the subsequent scan, shown in Fig. 1(e), resulting in a displacement of the upper portion of the white stripe, this time away from the trench. The final scan, taken immediately afterward and shown in Fig. 1(f), shows a larger portion of the white stripe is farther away from the trench, resulting in a structure clearly distinct from that in Fig. 1(b). These images help illustrate the size of the regions that can be impacted by an EM-STM measurement on graphite.

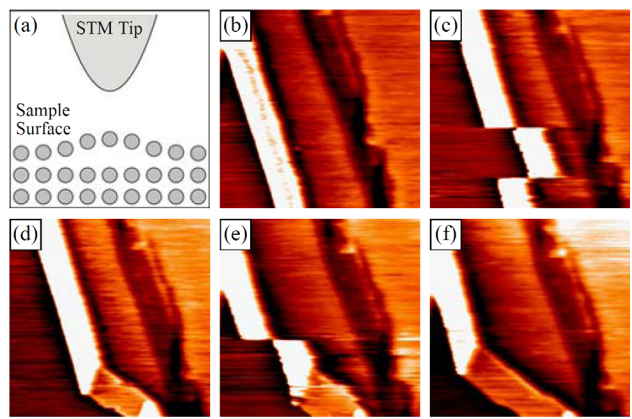

Fig. 1 – (a) A schematic of the STM tip lifting the surface layer of a graphite sample. (b-f) A chronological series of 150 nm × 150 nm filled-state STM images of one location on the graphite surface taken with a bias voltage of 0.1 V and a setpoint current of 1.0 nA. EM-STM measurements (not shown) were performed on the white stripe during the acquisition of the images shown in (c) and (e).

2.3. *EM-STM on pristine graphite terrace*

When EM-STM is carried out on a pristine flat terrace of graphite, the effect is different and is summarized in Fig. 2. First, a representative EM-STM measurement taken on graphite (solid line) and Au (dashed line) yields the height of the STM tip as a function of bias voltage as shown in Fig. 2(a). The measured tunneling current is also plotted in the inset to show that it remains at an approximately constant value of 0.20 nA throughout the duration of both measurements. The EM-STM measurement on the graphite surface shows that during the voltage sweep from 0.1 V to 0.6 V, the tip is held at its initial height with little variation. From 0.6 V to 0.7 V, the tip swiftly retracted by about 30 nm, at which height it roughly stabilized. This behavior is consistent with the idea that the top layer of graphite is held in place by the bulk until the electrostatic force of attraction, which increases with voltage, becomes large enough to locally separate it. The measured tunneling current serves as evidence that the sample surface must move with the tip. If it did not, the current would exponentially fall to zero around 0.6 V. Note that traditional constant-height (feedback off) STS data was also acquired (not shown), but the current quickly saturated the preamplifier, consistent with the sample crashing into the stationary STM tip. Our EM-STM data for graphite is compared with that for the bare Au surface, in which the tip height increased only slightly across the same voltage range. Thirty times larger displacements of the STM tip occur for EM-STM on HOPG than on Au.

## 3. Results and Discussion

The approximate force between the tip and the graphite as a function of bias voltage was calculated using the method of images [17]. The tip is modeled as a biased conducting sphere of radius 20 nm and the graphite is modeled as an infinite grounded conducting plane. The initial sphere-plane separation was set at 0.5 nm, but this value was adjusted as the voltage increased to correct for the small vertical movement observed in a stationary control sample of graphene on copper foil. The calculated force vs. voltage data was then combined with the experimental EM-STM data for HOPG in Fig. 2(a) to plot the attractive electrostatic force as a function of tip height in Fig. 2(b). This curve shows that the surface does not lift significantly until a load force of about 0.2 nN is applied, after which it is easy to raise (effective spring constant of ~2 pN/nm) for about 30 nm. The shaded region under the curve has an area of about 50 eV, corresponding to the energy expended to lift the surface layer.

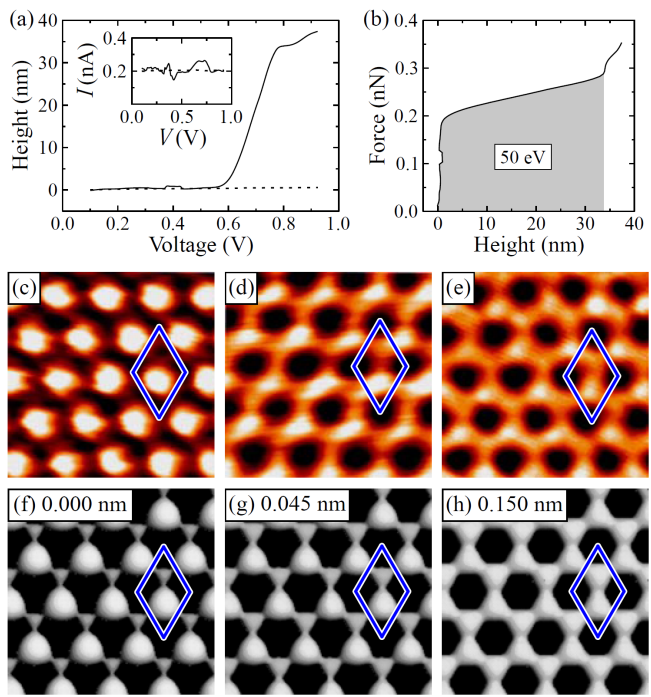

Fig. 2 – (a) The height of the STM tip as a function of the tip bias during an EM-STM measurement on HOPG (solid line) and on Au (dashed line). The measured tunneling current is plotted as a function of voltage for both in the inset. (b) Force exerted by the STM tip on the HOPG surface as a function of tip height, based on a method-of-images calculation. Shaded region indicates energy expended. (c-e) Filled-state atomic-resolution STM images of the HOPG surface taken with a bias voltage of 0.1 V and a setpoint current of 0.2 nA. A unit cell is superimposed on each image. Notice that (c) shows triangular symmetry because only alternate atoms appear in the image, while (e) shows the full hexagonal symmetry. (f-h) Simulated STM images of graphite taken from DFT calculations. The rhombus unit cell is again superimposed on each image, and the top layer's displacement from equilibrium is indicated at top left.

Next, three atomic-resolution STM images of the HOPG surface are presented in Fig. 2(c-e). Each possesses a different symmetry, highlighted by the rhombus-shaped unit cell superimposed on each image. A typical STM image of HOPG is shown in Fig. 2(c), with white spheres representing the B atoms arranged with trigonal symmetry. For this image, the unit cell depicts only one atom. The bright white features are still present in Fig. 2(d), but now the A atoms are also somewhat visible, resulting in an asymmetrical hexagonal pattern. Two atoms are now apparent in the unit cell, but with a larger charge density on the bottom atom. Finally, a more balanced hexagonal pattern is observed in Fig. 2(e). Both atoms in the unit cell possess nearly equal charge density, resembling a typical STM image of graphene rather than graphite. This type of image on HOPG is much less common than the first one, and in the past obtaining it has mostly been a matter of chance. However, EM-STM provides a mechanism for directly separating the surface layer from the bulk at will, effectively creating a section of graphene. By systematically repeating the EM-STM measurement at successively higher voltages, one can tune the displacement of the top layer. While this procedure does lift the layer, the top layer is still attracted to the graphite and thus quickly relaxes. Nevertheless, the likelihood of observing the graphene hexagonal symmetry on HOPG does greatly increase after repeatedly performing EM-STM.

A full understanding of our experimental findings was not possible until simulated STM images of HOPG were extracted from DFT calculations [12]. These calculations were performed within the local-density approximation to DFT, without modeling the STM tip [18] and using projector augmented-wave potentials [19] as implemented in the plane wave basis set VASP [20] code. The graphite was modeled as a six-layer Bernal stack, using a $1 \times 1$ unit cell. A cutoff energy of 500 eV and a very large $219 \times 219 \times 1$ Monkhorst-Park $k$-point mesh were used to ensure proper sampling around the Dirac point. Initially, the atoms were allowed to move until all forces were less than 0.1 eV/nm, resulting in a carbon-carbon bond length of 0.142 nm and an interplanar separation of 0.334 nm. Then the top layer was moved away from the bulk in ten steps of 0.015 nm, allowing only in-plane relaxation at each step. For each configuration, a simulated constant-current STM image was produced by integrating the local DOS from the Fermi level to 0.06 eV below that point and choosing an appropriate isocontour surface. These parameters were chosen to best replicate the experimental STM conditions.

Three simulated STM images taken from the DFT calculations are presented in Fig. 2(f-h). For each, the displacement of the top plane relative to its equilibrium position is listed in the top left corner. The first image displays large spheres representing the electron density around the B atoms arranged in a trigonal pattern as shown in Fig. 2(f). Smaller triangles represent the electron density around the A atoms, unresolved in the experimental STM images. After a vertical displacement of 0.045 nm, the circles have shrunk while the triangles have grown larger and more rounded. At 0.150 nm the electron density about each atom is essentially equivalent, with no significant changes occurring with further displacements. As can be seen by comparing the unit cells in corresponding figures, the simulated images are in excellent agreement with the experimental data.

More information about the electronic properties throughout the displacement can be found in the band structure at each step. A side view of the six-layer simulated structure after the the top layer has been displaced vertically by 0.090 nm is shown in Fig. 3(a). Notice how the charge density of the top layer is clearly separated from the bulk layers and more concentrated. The band structure properties near the K-point for the six-layer graphite structure (without any top layer displacement) are shown in Fig. 3(b). As expected all the bands are parabolic. The band structure after the top was displaced 0.150 nm now includes some linear behavior, which is characteristic of graphene as shown in Fig. 3(c). Note, there is an extra set of linear bands coming from the odd number of layers remaining in the split-off graphite structure [21]. After analysis of the band structure throughout the movement of the top layer, we estimate that around 0.090 nm the unique electronic properties of graphene are fully present. Namely, the bands near the K-point are linear and the total surface charge density has increased to nearly the level of isolated graphene. Next, the net energy change of the total graphite system is plotted versus the top layer's vertical displacement from equilibrium in Fig. 3(d). The displacement is reported as a percentage of the equilibrium interplanar separation (0.334 nm), or the uni-axial strain $\varepsilon_{zz}$. The energy curve increases smoothly over the range sampled, and it transitions from positive to negative curvature near a strain of 13.5% (or a displacement of 0.045 nm). This inflection point is identified with an arrow. The calculated energy needed to fully separate the unit cell is found to be approximately 50 meV. From our earlier estimates we found that the STM tip expended 50 eV lifting the top layer 30 nm. Thus, we can now estimate that about 1,000 unit cells were separated during the lift. If the graphene was simply vertically lifted, a circular region with a radius of about 10 nm would be affected. Since this is similar to the height of the lifted graphene, we believe that a much larger area may slide across the graphite surface.

Next, we can estimate the force required to separate the layers by taking the derivative of the energy curve in Fig. 3(d), according to the Hellmann-Feynman theorem. This force (or uniaxial stress $\sigma_{zz}$) is a result of the attractive force between the graphitic layers, which increases up to the inflection point in the energy and subsequently decreases as shown in Fig. 3(e). The peak force required to separate the (1x1) layers is around 0.07 nN. This is smaller than the estimated electrostatic force applied by the STM tip (0.2 nN), which is consistent with the tip being able to lift the layer.

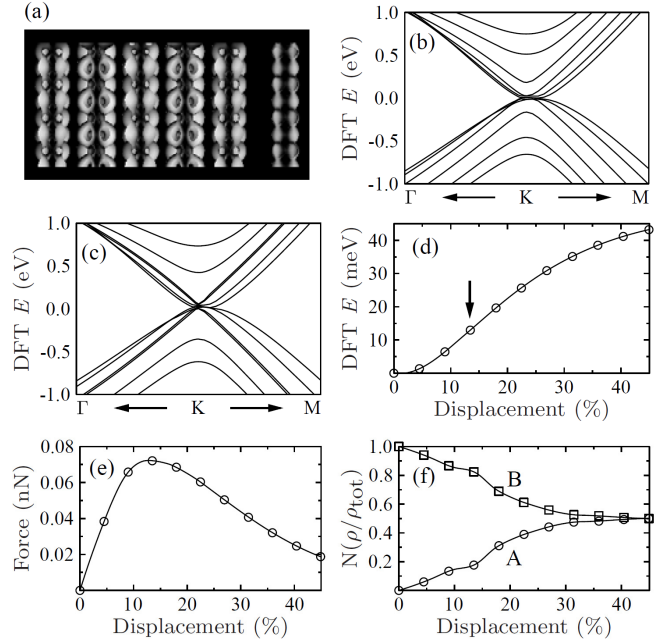

Fig. 3 – (a) Side view of the six-layer HOPG simulated structure shown with the top layer separated from the bulk by an additional 0.090 nm. (b) Band structure near the K-point for the six-layer HOPG structure without any movement of the top layer. (c) Band structure near the K-point for the six-layer HOPG structure after the top layer was vertically displaced by 0.150 nm away from the bulk. (d) DFT energy per unit cell as a function of displacement of the top layer of graphite. (e) The force as a function of displacement, obtained by taking the derivative of the energy with respect to displacement. (f) The normalized charge density on the A atom and the B atom as a function of displacement, taken from DFT simulated STM images.

Lastly, we present the charge density found on the A atom site ($\rho_A$) and the B atom site ($\rho_B$) as a function of layer separation in Fig. 3(f). These parameters have been normalized in two ways. First, since the total electronic charge in the top layer increased with the vertical displacement [12], every charge density was divided by the total charge density at that point, $\rho_{tot}$ = $\rho_A$ + $\rho_B$. This ensures that we track only changes in the relative charge densities ($\rho_A/\rho_{tot}$ and $\rho_B/\rho_{tot}$). Second, a normalization was applied to the data for each atom so that the normalized quantities, N($\rho_A/\rho_{tot}$) and N($\rho_B/\rho_{tot}$), vary from 0 to 0.5 and from 0.5 to 1, respectively. Thus, at zero displacement, N($\rho_A/\rho_{tot}$) is a minimum, and N($\rho_B/\rho_{tot}$) is a maximum, consistent with the STM images. Also, at the maximum displacement, the charge densities have equalized, also as seen in the STM images. (Note, these values are independent of the isovalue chosen for the simulated STM images.) A key benefit of this normalization scheme is that $N(\rho_B/\rho_{tot})$ represents a stepwise measurement of the decreasing interplanar coupling strength. If rescaled from 1 to 0, this parameter can be thought of as the effective mass scaling parameter [22]. The other parameter, $N(\rho_A/\rho_{tot})$, tracks the symmetry of the unit cell charge density. This parameter is tending toward zero as the symmetry between the A and B atoms is being broken. In this sense, this parameter (if rescaled from 0 to 1) represents the order parameter for the electronic reconstruction. The charge density profiles were also studied as a function of the bias voltage. For lower bias voltages (i.e., states closer to the Dirac point) the charge densities still began deviating from 50% at a strain around 40%, but the change to 1 or 0 happened more rapidly. This indicates that the states closer to the Fermi level are more sensitive to the surrounding environment.

In a broader context, we are modeling the case where a normal force is continuously applied to the graphene as it approaches graphite. The two systems eventually begin to interact, and the graphene transitions to a layer of graphite. Interestingly, if pressure were applied still further, a second transition would occur from graphite to diamond [23], as has been recently verified experimentally using femtosecond laser pulses to achieve the change [24]. However, what makes the graphene to graphite transition special is that it is the only known system where one can observe with atomic resolution how the electron acquires mass; or alternatively, how the electron loses mass and graphene generates its giant charge density responsible for its high current carrying capacity and thermal conductivity.

Our new EM-STM technique significantly broadens the abilities of the STM technique. STM is already known for its superior ability to obtain atomic structural and local electronic information for rigid samples. Now, if the sample is free to move or suspended, one can use EM-STM to gain insight into the local electrostatic and elastic properties [16]. This could prove valuable when considering chemically modified graphene, for example.

## 4. Conclusion

We have shown that EM-STM measurements can be used to reversibly and irreversibly alter an HOPG surface with considerable precision by varying the STM tip bias relative to the grounded sample. This technique was employed to physically alter the HOPG surface with precise spatial control. In addition, this technique was used to controllably lift the top HOPG layer away from the bulk. DFT simulated STM images for various displacements of the top layer relative to the bulk gave excellent agreement between the theoretical and experimental STM images. Band structure information predicts that the electronic properties of the top layer matched graphene after a displacement of 0.090 nm. Finally, by using the theoretical real-space charge densities to characterize the transition from graphite to graphene, a step-wise model of the interplanar coupling that is responsible for the electron acquiring effective mass was presented.

**Acknowledgements**

Special thanks are given to S. Barraza-Lopez and J. Tchakhalian for their insightful comments. P.X. and P.T. gratefully acknowledge the financial support of the Office of Naval Research (ONR) under grant number N00014-10-1-0181 and the National Science Foundation (NSF) under grant number DMR-0855358. Y.Y. and L.B. thank ARO Grant W911NF-12-1-0085, the Office of Basic Energy Sciences, under contract CR-46612 for personnel support and NSF grants DMR-0701558 and DMR-1066158; and ONR Grants: N00014-11-1-0384 and N00014-08-1-0915 for discussions with scientists sponsored by these grants. Calculations were made possible thanks to the MRI grant 0959124 from NSF, N00014-07-1-0825 (DURIP) from ONR, and a Challenge grant from HPCMO of the U.S. Department of Defense.